\documentclass{emulateapj}
\usepackage{amsmath,bm}
\usepackage{mathptmx}
\usepackage[]{units}
\usepackage{color}

\newcommand{\nuc}[2]{\ensuremath{\mathrm{^{#1}#2}}}
\newcommand{\ye}{\ensuremath{Y_e }}

\newcommand{\msun}{\ensuremath{\mathrm{M}_\odot}}

\newcommand{\nar}{New A Rev.}
\usepackage[tight]{subfigure}

\begin{document}
\shorttitle{Radioactive heating for SN~1987A light curve}
\shortauthors{Seitenzahl et al.}  \title{The light curve of SN~1987A
  revisited: \\ {constraining production masses of radioactive nuclides}} 
 \author{Ivo
  R. Seitenzahl\altaffilmark{1,2,3}, F.~X. Timmes\altaffilmark{4,5}, and
  Georgios Magkotsios\altaffilmark{5}}

\altaffiltext{1}{Research School of Astronomy and Astrophysics, Mount Stromlo Observatory,
Cotter Road, Weston Creek, ACT 2611, Australia} \altaffiltext{2}{Universit\"at W\"urzburg, Emil-Fischer-Stra{\ss}e 31,
  D-97074 W\"urzburg, Germany} \altaffiltext{3}{Max-Planck-Institut f\"ur
  Astrophysik, Karl-Schwarzschild-Stra{\ss}e 1, \\ D-85748 Garching,
  Germany} \altaffiltext{4}{School of Earth and Space Exploration,
  Arizona State University, Tempe, AZ 85287, USA} \altaffiltext{5}{The
  Joint Institute for Nuclear Astrophysics, Notre Dame, IN 46556, USA}
\email{ivo.seitenzahl@anu.edu.au}

\begin{abstract}
We revisit the evidence for the contribution of the long-lived
radioactive nuclides \nuc{44}{Ti}, \nuc{55}{Fe}, \nuc{56}{Co},
\nuc{57}{Co}, and \nuc{60}{Co} to the UVOIR light curve of SN~1987A.
We show that the V-band luminosity constitutes a roughly
constant fraction of the bolometric luminosity between 900 and 1900
days, and we obtain an approximate bolometric light curve out to
4334 days by scaling the late time V-band data by a constant factor 
where no bolometric light curve data is available.
Considering the five most relevant decay chains starting at
\nuc{44}{Ti}, \nuc{55}{Co}, \nuc{56}{Ni}, \nuc{57}{Ni}, and
\nuc{60}{Co}, we perform a least squares fit to the constructed composite
bolometric light curve. For the nickel isotopes, we obtain best fit values of 
M$(\nuc{56}{Ni}) = (7.1 \pm 0.3) \times 10^{-2}\, \msun$ and
M$(\nuc{57}{Ni}) = (4.1~\pm~1.8)\times10^{-3}\, \msun$.
Our best fit \nuc{44}{Ti} mass is M$(\nuc{44}{Ti}) = (0.55 \pm 0.17) \times 10^{-4}\, \msun$,
which is in disagreement with the much higher $(3.1 \pm 0.8) \times 10^{-4}\, \msun$ recently derived from \textsl{INTEGRAL} observations.
The associated uncertainties far exceed the best fit values for 
\nuc{55}{Co} and \nuc{60}{Co} and, as a result, we only give upper limits 
on the production masses of
M$(\nuc{55}{Co}) < 7.2 \times 10^{-3} \msun$ and
M$(\nuc{60}{Co}) < 1.7 \times 10^{-4} \msun$.
Furthermore, we find that the leptonic channels in the decay of
\nuc{57}{Co} (internal conversion and Auger electrons) are a
significant contribution and constitute up to 15.5\% of the total
luminosity. Consideration of the kinetic energy of these electrons is essential in 
lowering our best fit nickel isotope production ratio to $[\nuc{57}{Ni}/\nuc{56}{Ni}]~=~2.5\pm1.1$,
which is still somewhat high but is in agreement with gamma-ray observations and model predictions.
\end{abstract}

\keywords{nuclear reactions, nucleosynthesis, abundances ---
  supernovae: general --- supernovae: individual (SN~1987A)}

\section{Introduction}
\label{sec:intro}
The set of radioactive parent isotopes that have been used to model
the nuclear decay energy source terms for the light curve of SN~1987A
are \nuc{22}{Na}, \nuc{44}{Ti}, \nuc{56}{Ni}, \nuc{57}{Ni} and
\nuc{60}{Co} \citep[e.g.,][]{pinto1988a,woosley1989a,timmes1996a}.  It
is commonly held that \nuc{44}{Ti} decay plays a dominant role beyond
${\sim}2000$ days
\citep[e.g.,][]{lundqvist2001a,fransson2002a,motizuki2004a,jerkstrand2011a,larsson2011a}.
Recently, it was shown that heating by internal conversion and
Auger electrons emitted during the decay of \nuc{57}{Co} and Auger
electrons produced in the decay of \nuc{55}{Fe} can be the dominant
channels for the light curves of thermonuclear supernovae
\citep{seitenzahl2009d,seitenzahl2011b,roepke2012a}.  In this paper,
we re-evaluate the light curve of SN~1987A, taking into account these
previously neglected decay channels.

Several observational and theoretical efforts have estimated the mass of
radioactive \nuc{44}{Ti} synthesized in SN~1987A--see
Table~\ref{tab:1} for a compilation of results and references. 
Most of the derived observational estimates and model predictions do not agree within their
respective uncertainties.  
For example, modeling the strengths of
metal emission lines in the nebular phase leads to \nuc{44}{Ti} masses
$\gtrsim 1.0^{-4}\ \msun$, while upper limits derived from space-based
infrared spectroscopy generally find $\mathrm{few} \times 10^{-5}
\ \msun$ of \nuc{44}{Ti}.  Analysis of
$400\ \mathrm{ks}$ of {\em Chandra} ACIS data suggests a $2\sigma$ upper
limit of $\mathrm{M}(\nuc{44}{Ti}) < 2 \times 10^{-4}\ \msun$
\citep{leising2006a}, while analysis of $6\ \mathrm{Ms}$ of hard X-ray
data taken with the IBIS/ISGRI instrument on {\em INTEGRAL} suggests $(3.1
\pm 0.8) \times 10^{-4}\ \msun$ of \nuc{44}{Ti} \citep{grebenev2012a}.
Spherically symmetric hydrodynamic models of SN~1987A progenitors tend
to produce $\mathrm{few} \times 10^{-5}\ \msun$ of \nuc{44}{Ti},
explosions models with high energies and artificially imparted asymmetries in two dimensions 
appear to produce $\mathrm{few} \times 10^{-4}\ \msun$ of \nuc{44}{Ti} along
the poles of the model explosions \citep[e.g.,][]{nagataki_1997_aa,nagataki_1998_aa}, 
and efforts to model the SN~1987A light curve cluster around
$\mathrm{M}(\nuc{44}{Ti}) \approx 1.0^{-4}\ \msun$. At face value,
these model predictions for the mass of \nuc{44}{Ti} ejected are
smaller than allowed by the uncertainties of the {\em INTEGRAL}
measurement \citep{grebenev2012a}.

Since \nuc{56}{Ni} and \nuc{57}{Ni} are short lived, their mass ratio
M(\nuc{57}{Ni})/M(\nuc{56}{Ni}) is often expressed in units of the
corresponding ratio of the final decay products in the Sun
i.e.,\ \mbox{$[\nuc{57}{Ni} / \nuc{56}{Ni}] =
  \left[\mathrm{M}(\nuc{57}{Ni}) / \mathrm{M}(\nuc{56}{Ni})\right] /
  \left[\mathrm{M}(\nuc{57}{Fe}) /
    \mathrm{M}(\nuc{56}{Fe})\right]_\odot$}.  
The solar ratio is 
$\left[\mathrm{M}(\nuc{57}{Fe}) /
  \mathrm{M}(\nuc{56}{Fe})\right]_\odot \approx 0.0235$
\citep{cameron1982a, anders1989a, rosman1998a, lodders2003a,
  asplund2009a}.  
The mass ratio of \nuc{57}{Ni} to \nuc{56}{Ni} in SN~1987A, which is
prominently affecting light curve models between ${\sim}900 - 1800$ days, has
not reached consensus between the values inferred from observations
and light curve models (see Table~\ref{tab:1}).  For example, 
$[\nuc{57}{Ni} / \nuc{56}{Ni}] = 1.5 \pm 0.3^{stat} \pm 0.2^{sys}$ was derived from the gamma
ray flux of \nuc{57}{Co} with the OSSE instrument on the {\em Compton
Observatory} \citep{kurfess1992a}, upper limits of
M$(\nuc{57}{Co}) < 2.8 \times 10^{-3}\ \msun$ from the HEXE instrument
aboard MIR-KVANT also corresponds to roughly ratios of 1.5
\citep{sunyaev1991a}, and interpretations of ground-based infrared
spectroscopy also favor similar ratios.  Such moderate enhancement
ratios are also in agreement with spherically symmetric and asymmetric
hydrodynamic models of SN~1987A progenitors, which produce
$[\nuc{57}{Ni} / \nuc{56}{Ni}]$ between 0.5 and 2.5.  
On the other hand, values of $[\nuc{57}{Ni} / \nuc{56}{Ni}] \approx 5$
times solar were initially derived from light curve modeling
\citep{kumagai1991a,suntzeff1992a,dwek1992a}.  Such an enhanced ratio
was challenged by theoretical nucleosynthesis considerations
\citep{woosley1991a}, which limit $[\nuc{57}{Ni}/\nuc{56}{Ni}]$ to at
most four times solar and place the most likely value between 0.5 and
2.5 times solar (see also Section~\ref{sec:network}).  Subsequently, time-dependent
models that allow for the effects of non-equilibrium ionization have
been introduced as a solution to the overproduction problem.  Based on
such calculations, values for $[\nuc{57}{Ni}/\nuc{56}{Ni}]$ as low as two times solar have been claimed to be in agreement with the observations \citep{fransson1993a,fransson2002a}, although we maintain that the four times solar $[\nuc{57}{Ni}/\nuc{56}{Ni}]$ case in Figure 3 of \citet{fransson1993a} actually provides a much better fit to the data.

\citet{seitenzahl2011b} estimated that including the usually ignored 
contribution of Auger and internal conversion electrons of
\nuc{57}{Co} and \nuc{55}{Fe} in the heating budget might make
significant contributions to the light curve of SN~1987A. In this
paper, we refine that estimate.  In Section~\ref{sec:network}, we
discuss constraints on $[\nuc{57}{Ni}/\nuc{56}{Ni}]$ from standard,
parameterized post-explosion freeze-out profiles.  In
Section~\ref{sec:lc}, we present our analytic light curve model and 
discuss extending the observed UVOIR pseudo-bolometric light curves of
SN~1987A to later times; in Section~\ref{sec:results}, we present
the results of modeling the extended light curve.  We find that
(1) the $(3.1\pm0.8) \times 10^{-4}\ \msun$ of \nuc{44}{Ti} derived by \citet{grebenev2012a}
from gamma-ray observations with {\em INTEGRAL} are in conflict with our much lower value of
$M(\nuc{44}{Ti}) = (0.55 \pm 0.17) \times 10^{-4}\, \msun$
required to explain the luminosity of the late light curve;
(2) including the internal conversion and Auger electrons produced in
\nuc{57}{Co} decay reduces the mass of \nuc{57}{Ni} required for light
curve models; 
(3) within our uncertainties, \nuc{60}{Co} may be the dominant source of 
radioactive energy injection for a few years at intermediate times between.
In Section~\ref{sec:summary}, we conclude with a summary.

\begin{table*}
  \begin{center}
\caption{Measured, inferred, and predicted yield values for SN 1987A.   
  \label{tab:1} }
\begin{tabular}{cccclc} \tableline
  {$\mathrm{M}(\nuc{55}{Co})$\tablenotemark{a} [$10^{-3} \msun$]} & 
  {$\mathrm{M}(\nuc{60}{Co})$ [$10^{-4} \msun$]} & 
  {$[\nuc{57}{Ni} / \nuc{56}{Ni}]$} & 
  {$\mathrm{M}(\nuc{44}{Ti})$ [$10^{-4} \msun$]} & 
  {method}  &
  {References} \\  \tableline \tableline
  \nodata&\nodata&\nodata                          & $3.1\pm0.8$  & soft gamma-ray (IBIS/ISGRI, INTEGRAL)&1 \\
  $\lesssim 1$ & \nodata &\nodata                          & $<2.0$ ($2 \sigma$) & X-ray (ACIS, Chandra)&2 \\
  \nodata&\nodata&$< 1.5$                          & $ < 90 $      & X-ray(HEXE/Mir-Kvant)&3 \\
  \nodata&\nodata&$1.5\pm 0.3^{stat} \pm 0.2^{sys}$ & \nodata          & gamma-ray (OSSE/Compton GRO)&4 \\
  \nodata&\nodata&$1-2$                            & \nodata          & infrared spectroscopy(FIGS/AAT)&5 \\
  \nodata&\nodata&${\lesssim}1.5$                  & \nodata          & infrared spectroscopy(ESO)&6 \\ 
  \nodata&\nodata&\nodata                               & $<0.15$     & infrared spectroscopy (ISO/SWS)&7 \\ 
  \nodata&\nodata&\nodata                               & $<0.59$\tablenotemark{b}     & infrared spectroscopy (ISO/SWS)&8 \\ \hline
  \nodata&\nodata&\nodata                               & $4.00$      & nebular emission lines (HST)&9 \\ 
  \nodata&\nodata&\nodata                               & $1.0-2.0$   & nebular emission lines (HST)&10\\ 
  \nodata&\nodata&\nodata                               & $1.4\pm0.5$ & nebular emission lines (HST) &11\\ \hline
  \nodata&\nodata&$\gtrsim5$                       & $0.92$      & UVOIR light curve&12 \\ 
  \nodata&\nodata&$4.5\pm1.6$                      & $1.00$      & UVOIR light curve&13 \\ 
  \nodata&\nodata&$5\pm1$                          & $1.00$      & UVOIR light curve&14 \\
  \nodata&\nodata&${\sim}2$                        & $1.00$      & UVOIR light curve&15 \\
  \nodata&\nodata&${\sim}2$                        & $0.5-2.0$   & UVOIR light curve&16 \\ \hline
  \nodata&\nodata&$0.5-2.5$  & $\lesssim 0.85$      & pure nuclear reaction network calculation&17 \\
  \nodata&\nodata&${\sim}1.0$& $\lesssim 0.3$  & explosion model/network calculations&18 \\
  0.3    &3.6e-10&$1.7$      & $1.7$           & explosion model/network calculations&19 \\
  1.3    & 1.6   & 1.3       &  0.26  & S19 explosion model/network calculations&20 \\
  \nodata&\nodata&$0.5-2.0$  & $\lesssim 0.5$  & explosion model/network calculations&21 \\
  \nodata&\nodata&\nodata    & $\lesssim 0.5$  & explosion model/network calculations&22 \\
  {$\bm{<7.2}$} & {$\bm{<1.7$}} & {$\bm{2.5 \pm 1.1}$}  &  {$\bm{0.55 \pm 0.17}$} &  \textbf{least squares light curve fitting} & \textbf{this work} \\ \hline
\end{tabular}
\tablenotetext{1}{This is actually the combined mass of \nuc{55}{Co} and \nuc{55}{Fe}, but since most of the mass in the A=55 decay chain is synthesized as \nuc{55}{Co} and $\tau(\nuc{55}{Co}) << \tau(\nuc{55}{Fe})$ (see Section~\ref{sec:model}), we write the sum of the masses as $\mathrm{M}(\nuc{55}{Co})$.}
\tablenotetext{2}{Extreme assumptions about clumping and positron
  escape increase this upper limit to $<1.1$} 

\tablerefs{
  (1)~\citet{grebenev2012a}; (2)~\citet{leising2006a};
  (3)~\citet{sunyaev1991a}; (4)~\citet{kurfess1992a};
  (5)~\citet{varani1990a}; (6)~\citet{bouchet1993a};
  (7)~\citet{borkowski1997a}; (8)~\citet{lundqvist2001a};
  (9)~\citet{wang1996a}; (10)~\citet{chugai1997a};
  (11)~\citet{jerkstrand2011a}; (12)~\citet{kumagai1991a};
  (13)~\citet{dwek1992a}; (14)~\citet{suntzeff1992a};
  (15)~\citet{fransson1993a}; (16)~\citet{fransson2002a};
  (17)~\citet{woosley1991a}; (18)~\citet{woosley1995a};
  (19)~\citet{thielemann1996a}; (20)~\citet{rauscher2002a};
  (21)~\citet{nomoto2006a}; (22)~\citet{tur2010a}. }
\end{center}
\end{table*}

\section{Constraining $[\nuc{57}{N\lowercase{i}}/\nuc{56}{N\lowercase{i}}]$ 
from parameterized freeze-out profiles}
\label{sec:network}
\begin{figure}
  \includegraphics[width= 3.5in]{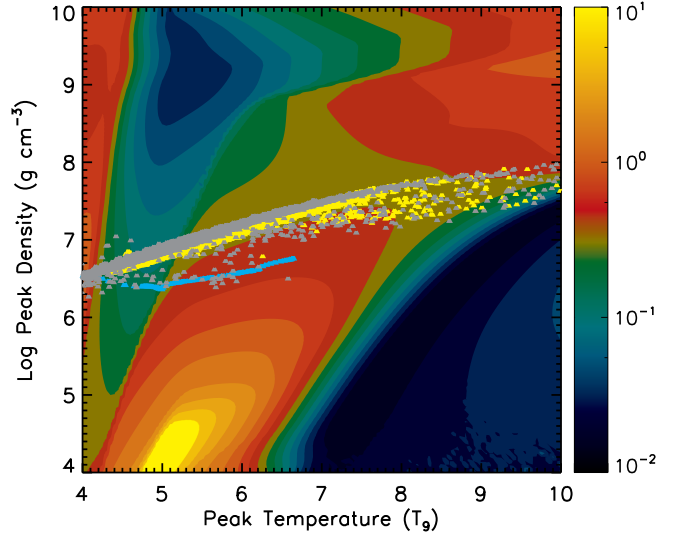}
  \caption{Final [\nuc{57}{Ni} / \nuc{56}{Ni}] after freeze-out from
    exponential thermodynamic trajectories for an electron fraction of
    $\ye=0.5$ as a function of peak temperature and peak density
    \citep{magkotsios2010a,magkotsios2011a}.  Over-plotted are the
    peak thermodynamic conditions reached in three multi-dimensional
    core-collapse supernova simulations of SN~1987A (cyan: P.~Young and C.~Fryer,
    private communication, 2012; yellow: A.~Wongwathanarat and
    T.~Janka, private communication, 2012; gray: \citealt{wongwathanarat2010a}).}
  \label{fig:1}
\end{figure}

Fig.\ \ref{fig:1} shows $[\nuc{57}{Ni} / \nuc{56}{Ni}]$ 
produced from post-explosion freeze-out expansion calculations for
exponential thermodynamic trajectories within a large grid of peak
temperatures and peak densities \citep{magkotsios2010a,magkotsios2011a}.  
Both \nuc{56}{Ni} and \nuc{57}{Ni} belong in the second family of
isotopes that are produced during freeze-out expansions \citep[see
  Table 2 of][]{magkotsios2011a}.  Isotopes of the second family
become nuclear flow hubs, dominate the final composition, and 
do not sustain any transition between equilibrium
states during the evolution.  This implies relatively featureless
contour plots of final yields compared to the isotopes of the first
family which do undergo a transition between equilibrium states during
the evolution.  The structure of $[\nuc{57}{Ni}/\nuc{56}{Ni}]$ in
\mbox{Fig.\ \ref{fig:1}} stems from a varying, relative efficiency between
\nuc{56}{Ni} and \nuc{57}{Ni} in absorbing nuclear flows.  Overall,
the doubly magic nucleus \nuc{56}{Ni} is the most efficient flow hub
among the isotopes of the second family near the magic number 28.  The
relative strength of \nuc{57}{Ni} to \nuc{56}{Ni} in absorbing flow
has a weak dependence on the type of freeze-out. In particular,
\nuc{57}{Ni} is not very efficient for the $\alpha p$-rich freeze-out
region and parts of the Si-rich and normal freeze-out regions. For the
photodisintegration regime at the bottom right part of
\mbox{Fig.\ \ref{fig:1}} neither isotope is produced.

Fig.~\ref{fig:1} also shows the peak conditions taken from three
multi-dimensional supernova simulations of SN~1987A to constrain the
accessible $[\nuc{57}{Ni} / \nuc{56}{Ni}]$.  \mbox{Fig.\ \ref{fig:1}}
suggests that in these simulations most of the parameterized
trajectories produce ratios near or below unity.  The total
$[\nuc{57}{Ni} / \nuc{56}{Ni}]$ ratio of all mass elements from the
exponential trajectories are generally are within a factor of $\sim$2
of the ratios found by post-processing the multi-dimensional SN~1987A
supernova simulations with the same 489 isotope reaction network used
for the parameterized trajectories.  These results are consistent 
with measurements in the 1.5 -- 2 range  listed in Table~\ref{tab:1}.

\section{The UVOIR light curve}
\label{sec:lc}

\begin{figure}
  \includegraphics[width= 3.5in]{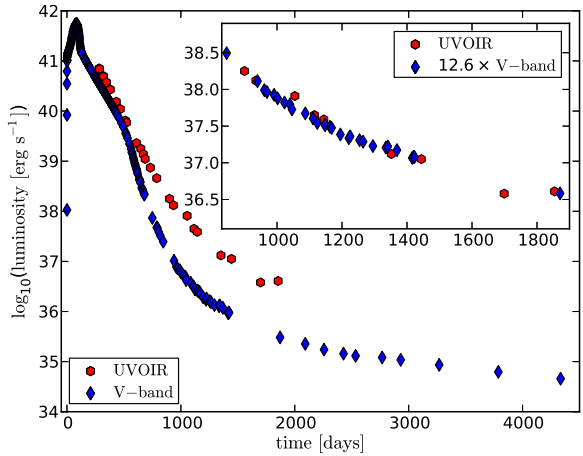}
  \caption{UVOIR \citep{suntzeff1992a,suntzeff1997a} and V-band
    \citep{leibundgut2003a,fransson2007a} light curves for SN~1987A.
    The inset shows that the V-band is approximately a constant factor of
    12.6 the bolometric light curve between 840 and 1854 days.  We use
    this agreement to approximate the UVOIR light from the V-band by
    assuming the same scaling behavior holds also between days 1854
    and 4436.}
  \label{fig:2}
\end{figure}

We base our analysis on the UVOIR pseudo-bolometric light curves
(i.e., not counting escaping gamma-rays) instead of a wavelength
dependent radiative transfer calculation.  \mbox{Fig.\ \ref{fig:2}} shows
V-band data to 4334 days \citep{leibundgut2003a,fransson2007a} and
UVOIR data to 1854 days \citep{suntzeff1992a,suntzeff1997a} for
SN~1987A.  Owing to their long half-lives, the radionuclides
\nuc{44}{Ti}, \nuc{55}{Fe}, and \nuc{60}{Co} mainly affect the 
bolometric light curve in the later phases after ${\sim}1200$ days.  
To constrain the mass ejected of these isotopes, we notice the V-band is an
approximately constant fraction of the bolometric light between 840
and 1854 days, as shown by the inset of \mbox{Fig.\ \ref{fig:2}}.  
We construct an approximate bolometric light curve for longer than 1854
days by scaling the V-band by a constant factor of 12.6. This provides
an additional ten UVOIR data points at late times that correspond to
the ten measured V-band data points shown in \mbox{Fig.\ \ref{fig:2}}.

\subsection{Light curve model}
\label{sec:model}
We model the bolometric light curve with effective opacities following
\citet{woosley1989a,timmes1996a}.  We assume that only radioactive energy
input contributes to the luminosity and approximately account for
energy stored in ionization of the ejecta and released at a later
time. We include energy deposited by Auger and internal conversion
electrons and we use up to date nuclear decay information from the
National Nuclear Data Center\footnote{http://www.nndc.bnl.gov/} (see
Table~\ref{tab:2}).  We consider contributions from the following five
decay chains to the bolometric supernova light curve:
\begin{align}
  & &^{56}\mathrm{Ni} \;\stackrel{t_{1/2} = \; 6.08d}{\hbox to
    60pt{\rightarrowfill}} \; ^{56}\mathrm{Co} \;
  \stackrel{t_{1/2} = \; 77.2d}{\hbox to 60pt{\rightarrowfill}} \; ^{56}\mathrm{Fe} \\
  & &^{57}\mathrm{Ni} \;\stackrel{t_{1/2} = \; 35.60 h}{\hbox to
    60pt{\rightarrowfill}}\; ^{57}\mathrm{Co} \;
  \stackrel{t_{1/2} = \; 271.79d}{\hbox to 60pt{\rightarrowfill}} \; ^{57}\mathrm{Fe}\\
  & &^{55}\mathrm{Co} \;\stackrel{t_{1/2} = \; 17.53 h}{\hbox to
    60pt{\rightarrowfill}}\; ^{55}\mathrm{Fe} \;
  \stackrel{t_{1/2} = \; 999.67 d}{\hbox to 60pt{\rightarrowfill}} \; ^{55}\mathrm{Mn}\\
  & &^{44}\mathrm{Ti} \;\stackrel{t_{1/2} = \; 58.9 y}{\hbox to
    60pt{\rightarrowfill}}\; ^{44}\mathrm{Sc} \; \stackrel{t_{1/2} =
    \; 3.97 h}{\hbox to 60pt{\rightarrowfill}} \; ^{44}\mathrm{Ca} \\
  & &^{60}\mathrm{Co} \; \stackrel{t_{1/2} = \; 5.27 y}{\hbox to
    60pt{\rightarrowfill}} \; ^{60}\mathrm{Ni}
\end{align}
We do not consider the \nuc{22}{Na} decay chain since nucleosynthesis
calculations show the \nuc{55}{Co} chain always injects significantly more
energy than the \nuc{22}{Na} chain, and \nuc{55}{Fe} and \nuc{22}{Na}
have very similar half-lives of ${\sim}2.7$ and ${\sim}2.6$ years
respectively.

The time-dependence of $n$ nuclide abundances $N_i$ in a decay chain
is governed by the Bateman equations:
\begin{eqnarray}
  \label{eq:bateman}
  \frac{dN_1}{dt}&=&-\lambda_1 N_1\\
  \frac{dN_i}{dt}&=& \lambda_{i-1} N_{i-1}-\lambda_{i} N_{i} \enskip .
\end{eqnarray}
For $n=2$ and initial abundances $N_1(0)$ and $N_2(0)$ we get the
solution
\begin{eqnarray}
  \label{eq:sol1}
  N_1(t)&=& N_1(0)\exp(-\lambda_1 t) \\
  \label{eq:sol2}
  N_2(t)&=& N_1(0) \frac{\lambda_1}{\lambda_2-\lambda_1} [
  \exp(-\lambda_1 t)-\exp(-\lambda_2 t) ] \\ \nonumber 
  & & + N_2(0)\exp(-\lambda_2 t) \enskip .
\end{eqnarray}
The decay constants $\lambda_i$ are related to the half-lives
$t_{1/2,i}$ and the mean life-time $\tau_i$ via
\begin{equation}
  \lambda_i = \frac{1}{\tau_i} = \frac{\ln(2)}{t_{1/2,i}} \enskip .
\end{equation}
The rate of energy deposition by decays of nucleus $i$ is given by the
activity multiplied by the energy deposited per decay:
\begin{equation}
  \epsilon_i = \lambda_i N_i(t) q_i(t) 
\end{equation}
where the number $N_i$ is given by eq.~\ref{eq:sol1} or
eq.~\ref{eq:sol2} and the energy deposited, $q_i$, is a function of
time due to the increasing escape fraction of gamma-rays and
possible late time escape of positrons.

To reduce the number of variables we make use of the large
difference in half-lives in four of the decay chains
\begin{eqnarray}
  \tau(\nuc{44}{Ti}) >> \tau(\nuc{44}{Sc})\\
  \tau(\nuc{55}{Co}) << \tau(\nuc{55}{Fe})\\
  \tau(\nuc{56}{Ni}) << \tau(\nuc{56}{Co})\\
  \tau(\nuc{57}{Ni}) << \tau(\nuc{57}{Co}).
\end{eqnarray}
This allows us to approximate the solution to the Bateman equations
with a single exponential for each decay chain. We refer to the
luminosity in the $A=44$ chain, where $A$ is the atomic number, as
being due to the long-lived \nuc{44}{Ti}, even though the positron is
actually produced in the subsequent decay of \nuc{44}{Sc}.  Using
effective opacities for the gamma-rays and hard X-rays and assuming
instantaneous and complete deposition of the leptonic kinetic energy,
we obtain the following time-dependent expression for the luminosity
due to a given decay chain

\begin{equation}
\label{eq:lum}
  L_{A}(t) = 2.221\ \frac{\lambda_{A}}{A} \frac{\mathrm{M}(A)}{\msun} 
  \frac{q^{l}_{A}+q^{\gamma}_{A} f_{A}}{\mathrm{keV}}
  \exp(-\lambda_{A} t) \times 10^{43} \mathrm{erg}\ ; 
\end{equation}

\noindent where $q^l$ and $q^{\gamma}$ are the average energies per decay
carried by charged leptons and gamma-rays respectively
(see~Table~\ref{tab:2}), \mbox{$f_{A}=\{1.0-\exp[-\kappa_{A} \phi_0(t_0/t)^2]\}$}, 
$\phi_0 = 7.0 \times 10^4$ g cm$^{-2}$ is the column density at the
fiducial time $t_0 = 10^6$ s, and $A$ stands for the atomic numbers
\{44,55,56,57,60\} of the five decay chains. 

Next, we describe how we take time-dependent freeze-out effects into 
account \citep[see][]{fransson1993a,fransson2002a}.
Freeze-out is most significant in the hydrogen envelope.  As
recombination times lengthen, the relative importance of freeze-out
effects initially increases, as long as gamma-rays still deposit
significant energy in the envelope. The \nuc{57}{Co} and \nuc{60}{Co} 
electrons, as well as the \nuc{44}{Sc} positrons are produced deep in
the core and have very short mean free paths. 
Complete and instantaneous thermalization is
therefore still a good approximation for these charged particle contributions.
Therefore, upon entering the positron/electron dominated phase, recombination
times in the hydrogen envelope become more or less irrelevant, which
leads to the disappearance of time-dependent recombination effects.
This behavior is clearly shown in figure 2 of \citet{fransson2002a},
in which time-dependent effects begin to make a difference starting at
$\sim$1000 days and increase in importance until $\sim$1600 days. From
then on, the difference between the time-dependent and the steady
state calculations decreases steadily until $\sim$2800 days when the
time-dependent and the steady state results converge again.  

We extract the freeze-out contribution $L_{freeze}(t)$ with a plot digitizer from figure 1 of \citet{fransson1993a} 
by taking the difference between the two curves labelled ``\nuc{56}{Ni}-only'' and extrapolating after 2000 days. 
We assume a linear dependence of the freeze-out correction on 
the initial \nuc{56}{Ni} mass, i.e.\ the term that is added to the bolometric luminosity is $\tfrac{\mathrm{M}(\nuc{56}{Ni})}{0.07\,\msun}L_{freeze}(t)$. 
Such a freeze-out correction scaling linearly with M$(\nuc{56}{Ni})$ appears to be a good approximation
since addition of e.g. two times solar \nuc{57}{Co} does not significantly effect the freeze-out contribution (see figure 1 of \citealt{fransson1993a}). In other words, most of the freeze-out luminosity is energy stored in ionization from the \nuc{56}{Co} dominated phase, which justifies our approach.

\begin{table}
  \centering
  \caption{Charged lepton ($q^l$) and gamma-ray ($q^\gamma$) partial
    radioactive decay energies per decay, effective opacities ($\kappa$),
    and decay constants ($\lambda$).   \label{tab:2} }
  \begin{tabular}{ccrrc} \hline
    {Nucleus} & 
    {$\lambda$} & 
    {$q^{l}$} & 
    {$q^{\gamma}$} & 
    {$\kappa$} \\ 
    & [$\mathrm{d}^{-1}$] & [keV] & [keV] & [$\mathrm{cm}^{2} \mathrm{g}^{-1}$]\\ \hline
    \nuc{60}{Co} &3.600e-4&96.41   &2504  &0.04\footnote{We use the same value as \citet{timmes1996a}.}\\
    \nuc{57}{Co} &2.551e-3&17.82  &121.6 &0.0792\\
    \nuc{56}{Co} &8.975e-3&119.4 &3606  &0.033\\
    \nuc{55}{Fe} &6.916e-4&3.973   &1.635\footnote{Counting X-rays.}&\nodata\\
    \nuc{44}{Ti} &3.222e-5&596.0  &2275  &0.04\\ \hline
  \end{tabular}
\end{table}

\section{Confronting the light curve data with observations and theory}
\label{sec:results}
In the previous section, we have constructed a bolometric light curve
of SN~1987A from published bolometric data points where available (out
to 1854 days) and scaled V-band light at later times.
In the following, we use this composite light curve to constrain production masses of 
radionuclides. 
\begin{figure}
  \includegraphics[width= 3.5in]{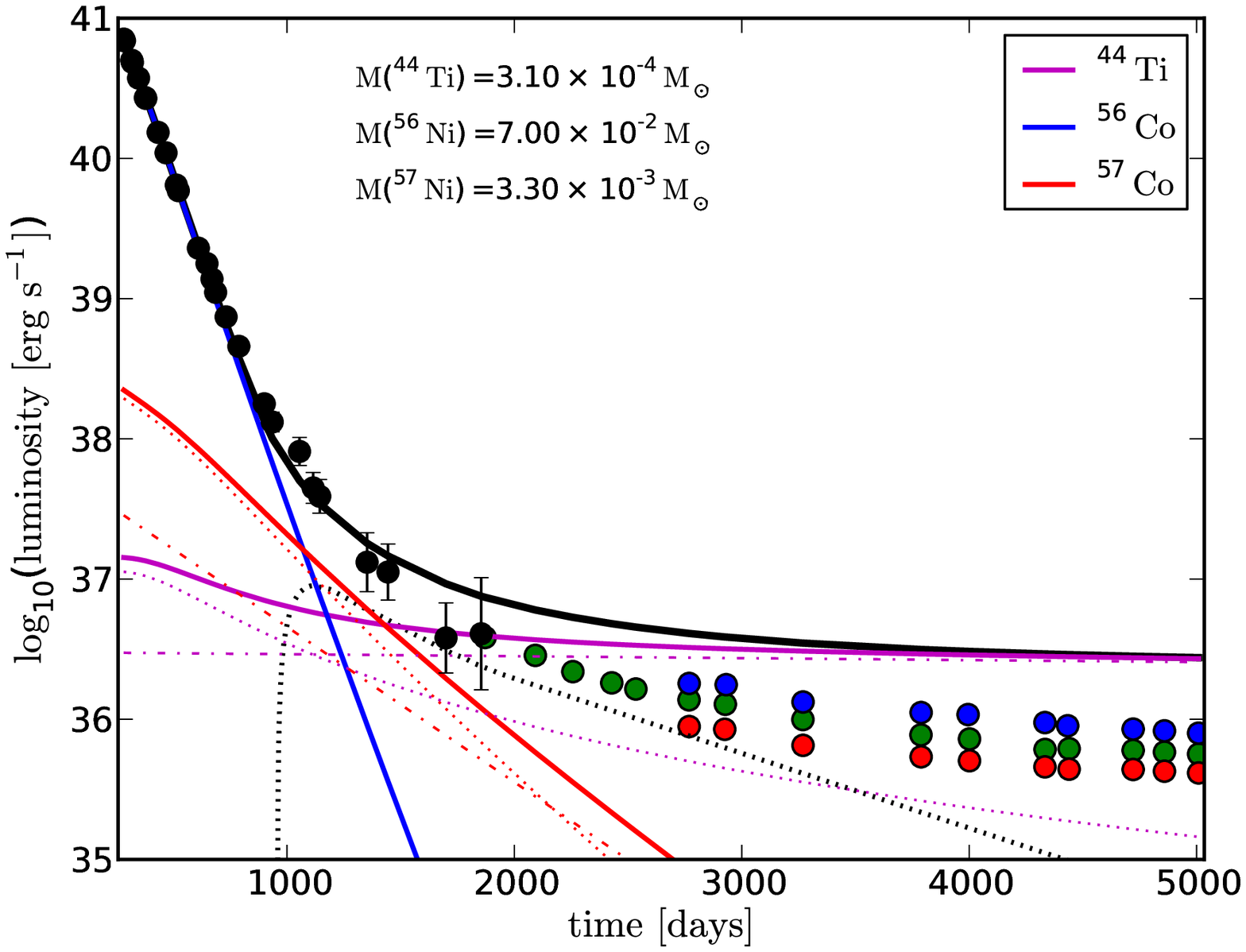}
\caption{Model light curve for canonically accepted values of \nuc{56}{Ni} and \nuc{57}{Ni} \citep{fransson1993a} and the \nuc{44}{Ti} value of \citet{grebenev2012a}. The partial $\gamma$-ray contributions for each nuclide are shown with dotted lines and electron and positron contributions are shown with dot-dashed lines. The solid black line is the model bolometric UVOIR luminosity when the energy stored in ionization and released later through recombination is approximately taken into account. The dotted black line shows this freeze-out term, i.e.\ our model for the delayed release of energy stored in ionization. In black are published UVOIR data for SN~1987A \citep{suntzeff1992a,suntzeff1997a}. In green are scaled V-band data \citep{leibundgut2003a,fransson2007a}. Arbitrarily scaled B-band (blue) and R-band (red) data extracted from the paper of \citet{larsson2011a} are shown to demonstrate that a similar rate of decline is present in other optical filters.} 
  \label{fig:3}
\end{figure}

We begin with a comparison of our constructed UVOIR light curve with the predictions of our light curve model for
canonical production masses of the radionuclides \nuc{56}{Ni}, \nuc{57}{Ni}, and \nuc{44}{Ti}. 
Fig.\ \ref{fig:3} shows the model UVOIR light curve and the luminosity of
the decay chains using canonical masses of 
$\mathrm{M}(\nuc{56}{Ni}) = 7.0 \times 10^{-2}\ \msun$ and 
$\mathrm{M}(\nuc{57}{Ni}) = 3.3 \times 10^{-3}\ \msun$
\citep{fransson2002a} and the recently determined
$\mathrm{M}(\nuc{44}{Ti}) = 3.1 \times 10^{-4}\ \msun$
\citep{grebenev2012a}. 
Green circles in \mbox{Fig.\ \ref{fig:3}} are the scaled and time extended V-band data where
the scaling corresponds to a constant fraction of $\sim$8\% (factor
12.6).  
\mbox{Fig.\ \ref{fig:3}} also shows arbitrarily
scaled R-band (red circles) and B-band (blue circles) data we extracted from figure 2 of
\citet{larsson2011a}.
Note that B-, V-, and R-band evolve quite similarly: there is apparently not much color evolution during these later epochs.
The fact that R- and B-band fall off in a similar manner and bracket the V-band in
wavelength lends credence to our extrapolation of the V-band scaling performed in Sec.~\ref{sec:lc}.
Note that all color light curves fall off with a time scale much faster than 
\nuc{44}{Ti} and that $\mathrm{M}(\nuc{44}{Ti}) = 3.1 \times 10^{-4}\ \msun$ results in a luminosity at late times significantly exceeding 
our composite light curve.

\begin{figure}
  \includegraphics[width= 3.5in]{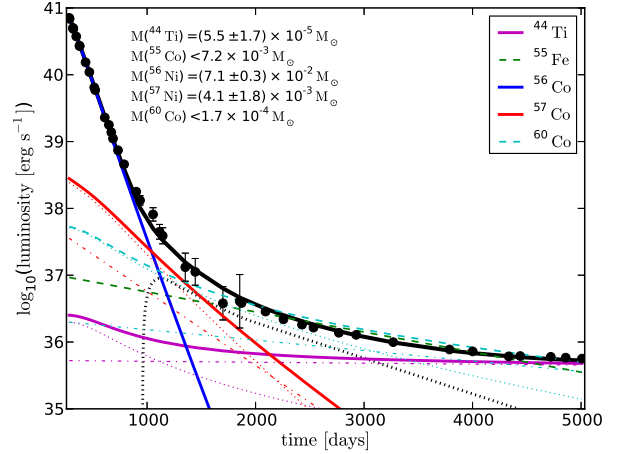}
\caption{Model light curve (thick black line) including our time-dependent freeze-out correction (black dashed line). This light curve is the result of a five component least squares fit of initial abundances of \nuc{44}{Ti}, \nuc{56}{Ni}, \nuc{57}{Ni}, \nuc{55}{Co} and \nuc{60}{Co} on the composite bolometric light curve constructed in Sec.~\ref{sec:lc}. The partial $\gamma$-ray contributions for each nuclide are shown with dotted lines and electron and positron contributions are shown with dot-dashed lines. The best fit values of \nuc{55}{Co} and \nuc{60}{Co} are significantly smaller than their respective uncertainties, which we show as upper limits for these isotopes.} 
  \label{fig:4}
\end{figure}

Next, we perform a non-linear, non-weighted least-squares fit
to the logarithm of composite light curve constructed in the previous
section.  We use {\sc SciPy} {\it curvefit}, which employs the
Levenberg-Marquardt (LM) algorithm.
We fit the data with the light curve model consisting of the
three traditionally employed radionuclides \nuc{56}{Ni}, \nuc{57}{Ni},
and \nuc{44}{Ti}, as well as \nuc{55}{Co} and \nuc{60}{Co} and the freeze-out
term (see Sec.~\ref{sec:model}). The result of the fit is shown in Fig.~\ref{fig:4}.  
For the nickel isotopes, we obtain best fit values of 
M$(\nuc{56}{Ni}) = (7.1~\pm~0.3)\times10^{-2}\, \msun$ and
M$(\nuc{57}{Ni}) = (4.1~\pm~1.8)\times10^{-3}\, \msun$.
Note that we obtain 
M$(\nuc{44}{Ti}) = (5.5 \pm 1.7) \times 10^{-5}\ \msun$, a value much smaller than the
{\em INTEGRAL} measurement of $(3.1
\pm 0.8) \times 10^{-4}\ \msun$  \citep{grebenev2012a},
but very similar to what is obtained from explosion models and nuclear reaction network calculations  (see Table~\ref{tab:1}).

The half-lives of \nuc{60}{Co} and \nuc{55}{Fe} are
quite similar, which introduces a degeneracy for the fitting algorithm.
As a result, their best fit values are much smaller than the associated uncertainties. 
While we use the best fit values of 
M$(\nuc{55}{Co})= 9.2 \times 10^{-6} \msun$ and 
M$(\nuc{60}{Co})= 4.5 \times 10^{-8} \msun$ for our combined light curve in Fig.~\ref{fig:4}, we 
plot the nuclide specific heating terms for the upper limits on the relevant production masses of
M$(\nuc{55}{Co}) < 7.2 \times 10^{-3} \msun$ and
M$(\nuc{60}{Co}) < 1.7 \times 10^{-4} \msun$. 

\begin{figure}
  \includegraphics[width= 3.5in]{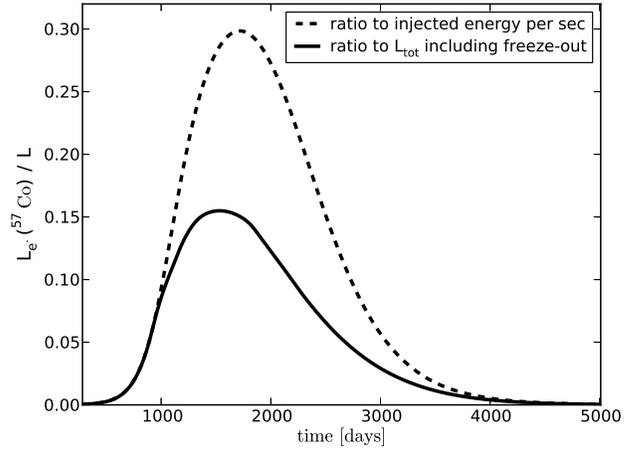}
\caption{Relative importance of the \nuc{57}{Co} electrons for the light curve of SN~1987A. The solid line is the ratio of the energy injection from Auger and internal conversion electrons of \nuc{57}{Co} (thin red dot-dashed line in Fig.~\ref{fig:4}) to the total luminosity including freeze-out for our best-fit solution shown in Fig.~\ref{fig:4} (solid black line). The dashed line is the ratio of the energy injection from Auger and internal conversion electrons of \nuc{57}{Co} to the instantaneous total rate of energy injection from radioactive decay (solid black line minus dotted black line in Fig.~\ref{fig:4}).}
  \label{fig:5}
\end{figure}

For SN~1987A, the leptonic channels of \nuc{57}{Co}
(thin red dash-dotted line) play a lesser role compared to the case of
thermonuclear supernovae. Nevertheless, their instantaneous, relative contribution
 to the total bolometric luminosity peaks at 1533 days at 15.5\% and constitutes over 10\% 
between 1058 and 2189 days (see Fig.~\ref{fig:5}). It is interesting to note that the effect 
of these electron channels on the best fit $[\nuc{57}{Ni}/\nuc{56}{Ni}]$ is even larger. If we 
perform a fit to the light curve with the electron channels of \nuc{57}{Co} omitted, we obtain 
$[\nuc{57}{Ni}/\nuc{56}{Ni}]~=~3.8\pm1.0$, whereas our best fit of the light curve that includes
the heating from internal conversion and Auger electrons of \nuc{57}{Co} yields 
$[\nuc{57}{Ni}/\nuc{56}{Ni}]~=~2.5\pm1.1$.

\begin{figure}
  \includegraphics[width= 3.5in]{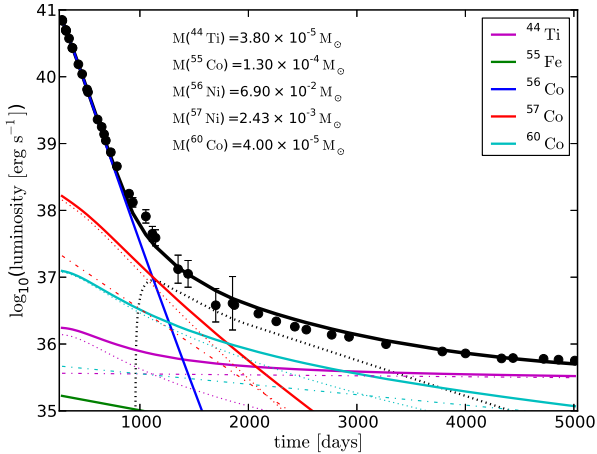}
\caption{Model light curve (thick black line) including our time-dependent freeze-out correction (dashed black line) for a choice of initial masses that results in a good fit  to the data. For \nuc{44}{Ti}, \nuc{57}{Co}, and \nuc{60}{Co}, gamma-ray (thin dotted lines) and leptonic (thin dash-dotted lines) partial light curves are shown separately as in Figs.~\ref{fig:3} and \ref{fig:4}.}
  \label{fig:6}
\end{figure}

Fig.~\ref{fig:6} demonstrates that a qualitatively good match to the observed light curve is possible where \nuc{60}{Co} dominates the energy injection in the epoch between the \nuc{57}{Co} and \nuc{44}{Ti} dominated phases. Such a scenario is a viable possibility and we advocate that \nuc{60}{Co} should be included in all late light curve and spectral models of SN 1987A. Again, we have taken the time-dependent effects of freeze-out in the hydrogen envelope approximately into account. This hypothetical model demonstrates that a good match to the light curve data of SN~1987A can be obtained with radionuclides other than \nuc{44}{Ti} (i.e.\ \nuc{60}{Co}) dominating the heating at intermediate times. 

The luminosity of X-rays produced in the radioactive decays is
generally very small compared to the other radiation from decays (e.g.\
positrons, electrons, gamma-rays) and neglecting X-rays for the
heating is therefore generally a good approximation. This ansatz
has been made in all previous works that fitted the light curve of SN
1987A.  We note, however, that since the decay of \nuc{55}{Fe} is a
ground state to ground state transition, no gamma-rays are emitted and
X-rays constitute $\sim$29\% of the total liberated decay energy
($\sim$41\% of the leptonic contribution).  The contribution of
X-rays is therefore quite significant for this particular nucleus and
for simplicity we work in the limit of full trapping and local
thermalization of \nuc{55}{Fe} X-rays. 

\section{Summary}
\label{sec:summary}
We have confirmed previous claims \citep{seitenzahl2011b} that the
leptonic heating channels (internal conversion and Auger electrons)
from the decay of \nuc{57}{Co} are a significant source of energy for
the light curve of SN~1987A, constituting as much as 30\% of the 
instantaneous total rate of energy injection from radioactive decay. For our best-fitting
abundances of the radionuclides, the \nuc{57}{Co} electrons constitute up to 15.5\%
of the total luminosity at 1533 days (see \mbox{Fig.\ \ref{fig:5}}).
Our best fit of the light curve that includes the heating from internal conversion and Auger electrons of 
\nuc{57}{Co} yields $[\nuc{57}{Ni}/\nuc{56}{Ni}]~=~2.5\pm1.1$, down from
$[\nuc{57}{Ni}/\nuc{56}{Ni}]~=~3.8\pm1.0$ for the case where these channels 
are omitted.

Here we only give upper limits M$(\nuc{55}{Co}) < 7.2 \times 10^{-3} \msun$ and M$(\nuc{60}{Co}) < 1.7 \times 10^{-4} \msun$ on the production masses of the cobalt isotopes. Within the uncertainties of our light curve analysis, an e.g. \nuc{60}{Co} dominated phase remains a possibility in the epoch between the \nuc{57}{Co}- and \nuc{44}{Ti}-dominated phases. We note, however, that our best fit to the reconstructed bolometric light curve (including the freeze-out term) does not require any significant power from either \nuc{55}{Fe} or \nuc{60}{Co}.

We investigated the impact of the recently determined \nuc{44}{Ti}
mass of $3.1 \times 10^{-4}\ \msun$. We find that published, fully
reconstructed data is still reconcilable with this high value, but
that V-, B-, and R-band at late times indicate that a much lower value
would match the light curve better. Indeed, our fit of the constructed bolometric
light curve gives M$(\nuc{44}{Ti}) = (0.55 \pm 0.17) \times 10^{-4}\, \msun$, a value
in good agreement with most explosion models and nuclear reaction network calculations.

\acknowledgements 
We thank Annop Wongwathanarat, Thomas Janka, Chris Fryer, and Patrick
Young for making their core collapse supernova simulation data
available to us.  We also thank Josefin Larsson, Roland Diehl, 
and the anonymous referee for their helpful comments. I.R.S was supported by the ARC Laureate Grant FL0992131, the Deutsche
Forschungsgemeinschaft via the Emmy Noether Program (RO 3676/1-1) and
the GRK-1147. G.M. was supported by the NSF under Grant PHY 08-22648
for the Frontier Center ``Joint Institute for Nuclear Astrophysics"
(JINA). F.X.T. was partially supported by NSF AST1107484.

\end{document}